# Data-Driven Nonparametric Existence and Association Problems

Yixian Liu, Yingbin Liang, Shuguang Cui

*Abstract*—We investigate two closely related nonparametric hypothesis testing problems. In the first problem (i.e., the existence problem), we test whether a testing data stream is generated by one of a set of composite distributions. In the second problem (i.e., the association problem), we test which one of the multiple distributions generates a testing data stream. We assume that some distributions in the set are unknown with only training sequences generated by the corresponding distributions are available. For both problems, we construct the generalized likelihood (GL) tests, and characterize the error exponents of the maximum error probabilities. For the existence problem, we show that the error exponent is mainly captured by the Chernoff information between the set of composite distributions and alternative distributions. For the association problem, we show that the error exponent is captured by the minimum Chernoff information between each pair of distributions as well as the KL divergences between the approximated distributions (via training sequences) and the true distributions. We also show that the ratio between the lengths of training and testing sequences plays an important role in determining the error decay rate.

## I. INTRODUCTION

**W**E consider two closely related nonparametric hypothesis testing problems. We assume that there are a set of $M$ distinct discrete distributions $p_1, ..., p_M$, and a training data stream that consists of data samples drawn from each distribution is available if the corresponding distribution is unknown. Furthermore, a testing data stream is observed, which consists of $n$ samples drawn from an unknown distribution. The goal is to solve the following two problems: (1) whether the testing data stream is drawn from one of the $M$ distributions; and (2) if the answer is yes, then which distribution generates the testing data stream. Clearly, the first problem tests the *existence*, i.e., whether the data stream belongs to a set of existent distributions, which is a binary composite hypothesis testing problem. The second problem tests the *association*, i.e., which distribution the data stream is associated with.

For parametric scenarios, where all distributions $p_1, ..., p_M$ are known in advance, both the existence and association problems have been well studied, e.g., [1], [2]. For nonparametric scenarios, where the distributions are unknown, but instead, a training data stream generated from each distribution is given, previous studies [3]–[5] focused only on the Neyman-Pearson formulation, i.e., given the requirement on the error probability for some hypotheses, the error probability for the remaining hypothesis needs to be minimized. The focus of this paper is to solve these problems in the nonparametric case based on a different performance criterion: the maximum of all types of error probabilities. Our focus is on the characterization of the error exponents for such error performance metrics as the sample size enlarges. Our study suggests that such a different performance criterion offers very different understanding and insights about these two problems.

The existence and association problems naturally correspond to two stages of detection problems in many practical applications. For example, in order to detect the operation mode in a cognitive radio (CR) system, a number of CR operation traces are initially collected in the profile, which represent the normal operational patterns of CR systems. Then given an observed CR trace, the goal of malware detection is to first determine whether the test CR trace belongs to the existing profile, i.e., whether it is normal or anomalous. If it is anomalous, an alarm of malware infection is triggered. Otherwise, the system needs to determine which existing class the observed operation trace is from, in order to track the operation of the system. The problems considered here can also be used in other applications such as speaker voice testing [6], [7], signal source and channel detection in wireless networks [8], and homogeneity testing and classification [9], [10].

### A. Contributions

In this paper, we construct generalized likelihood (GL) tests for the two nonparametric hypothesis testing problems, and characterize the error exponents for the tests. For the existence problem, we show that the error exponent is mainly captured by the Chernoff information between the set of composite distributions and alternative distributions. For the association problem, we show that the error exponent is captured by the minimum Chernoff information between each pair of distributions as well as the KL divergences between the approximated distributions (via training sequences) and the true distributions.

The work of Y. Liu was supported by University of Chinese Academy of Sciences under UCAS Joint PhD Training Program UCAS[2015]37. The work of Y. Liang was supported in part by DARPA FunLoL program and by NSF grants ECCS-1609916 and CCF-1617789. The work of S. Cui was supported in part by DoD with grant HDTRA1-13-1-0029, by grant NSFC-61328102/61629101, and by NSF with grants DMS-1622433, AST-1547436, ECCS-1508051/1659025, and CNS-1343155.

Y. Liu is with the Shanghai Institute of Microsystem & Information Technology, Chinese Academy of Sciences, Shanghai, China and also with the University of Chinese Academy of Sciences, Beijing, China and also with the School of Information Science & Technology, ShanghaiTech University, Shanghai, China (email: liuyx@shanghaitech.edu.cn).

Y. Liang is with the Department of Electrical and Computer Engineering, The Ohio State University, Columbus, OH, USA (email: liang.889@osu.edu).

S. Cui is with the Department of Electrical and Computer Engineering, University of California, Davis, CA, USA (e-mails: sgcui@ucdavis.edu).

We also show that the ratio $\beta$ between the lengths of training and testing sequences plays an important role in determining the error decay behavior. If $\beta \to \infty$, i.e., the length of training sequences scales order-wise faster than that of the testing sequence, then the error exponents of the considered nonparametric models converge to those of the parametric problems, and hence are optimal. If $0 < \beta < \infty$, i.e., the lengths of training and testing sequences scale at the same rate, then the GL tests are exponentially consistent (i.e., the error exponents are positive). Finally, if $\beta \to 0$, i.e., the length of training sequences scales order-wise slower than that of the testing sequence, then the tests are not exponentially consistent.

*B. Related Work*

The multiple hypothesis testing problems have been extensively studied, e.g., [1] and [2]. For the parametric case, the likelihood ratio test has been studied, and the optimal exponent of the error probability has been characterized in [2]. For the nonparametric case, most studies in the literature, e.g., [3]–[5], focused on the Neyman-Pearson formulation. Our work here on nonparametric hypothesis testing analyzes the exponent of the maximum error probability to quantify the performance.

The binary composite hypothesis testing problem has also been studied, e.g., in [11]–[13]. All of these studies focued on the parametric model and adopt the Neyman-Pearson formulation. We study the nonparametric binary composite hypothesis testing in this paper and prove that the nonparametric test is exponentially consistent in terms of the maximum error probability if the length of training sequences is large enough.

Our problem is also related to but different from the following models recently studied. One type of anomalous sample detection problems was studied in [14]–[16], in which given a training set of samples generated by one (or more) typical distributions, a new sample needs to be tested whether it is generated from the typical distributions or from an anomalous distribution (for example, from a mixture distribution of the typical distribution and another different distribution). Our model has a sequence of testing samples and our study focuses on characterizing the error exponent, whereas the previous studies did not analyze the error exponent. Furthermore, our existence problem can be viewed as a generalization of the two sample problem, in which two sets of samples respectively generated from two unknown distributions are available, and the goal is to test whether the two underlying distributions are identical. Our study is different from the previous studies in [17]–[20] in that our focus is on discrete distributions and characterization of the error exponent, whereas previous studies did not characterize the error exponent.

Our problem is also related but different from a class of outlying sequence detection problems studied in [21]–[24]. Our model has training sequences associated with hypotheses, whereas the previous results did not consider such information. The problem on testing closeness of distributions was studied in [25], [26], in which two sets of samples generated from two unknown discrete distributions are available, and the goal is to test whether the two distributions are close in the $\ell_1$ norm using as few samples as possible. Our existence problem is similar but has composite distributions, and the performance metric of the error exponent is also different from the previous studies.

*C. Organization*

The rest of the paper is organized as follows. In Section 2, we describe our problem formulation. In Section 3, we provide our main results on characterization of the error exponents for the two problems. In Section 4, we provide numerical results, and finally in Section 5, we conclude the paper.

## II. Problem Formulation

In this section, we formally define the existence and association problems that we study in this paper.

Suppose there are in total $M$ distinct discrete distributions $p_1, ..., p_M$ with the support set $\mathcal{Y}$. We consider a general nonparametric model, in which $p_i$'s for $1 \leqslant i \leqslant M_1$ are known, and $p_i$'s for $M_1 < i \leqslant M$ are unknown, where $0 < M_1 \leqslant M$. Clearly, if $M_1 = M$, the problem is fully parametric with all distributions known. If $M_1 = 0$, the problem is fully nonparametric with all distributions unknown. Thus, the model we study here unifies the parametric, semiparametric and nonparametric models; but we refer to such a model as a nonparametric model for simplicity, to which we make our main technical contributions. For each unknown distribution $p_i$ with $i > M_1$, a training sequence $t_i$ is available, which consists of $\bar{n}$ i.i.d. samples generated by $p_i$. Furthermore, a testing sequence $y$ is observed, which consists of $n$ i.i.d. samples generated by a certain distribution $p_s$ over the same support $\mathcal{Y}$. We assume that if $p_s \notin \{p_1, \ldots, p_M\}$, then $p_s$ is at least a certain distance away from these distributions. More formally, we require that

$$\min_{1 \leqslant i \leqslant M} C(p_s, p_i) \geqslant \alpha, \quad \text{for some } \alpha > 0, \quad (1)$$

where $C(p, q)$ is the Chernoff information [2] defined as

$$C(p, q) = \max_{\lambda \in [0,1]} -\log\left(\sum_{y \in \mathcal{Y}} p(y)^\lambda q(y)^{1-\lambda}\right). \quad (2)$$

We first consider the following two hypotheses:

$$H_0: p_s \notin \{p_1, \ldots, p_M\};$$
$$H_1: p_s \in \{p_1, \ldots, p_M\}.$$

The goal of our first problem is to determine which hypothesis occurs, i.e., whether the testing sequence is generated by one of $M$ existent distributions. We refer to this problem as the *existence problem*. Since the hypothesis $H_1$ contains $M$ sub-hypotheses, the existence problem is a binary composite hypothesis testing problem.

We let $\delta(y)$ denote a test rule for the existence problem, which maps the testing sequence $y$ to the corresponding hypothesis. Then for the existence problem, we take the following maximum of two types of error probabilities as the performance metric:

$$P_e(\delta) = \max\{P(\delta = 0|H_1), P(\delta = 1|H_0)\}. \quad (3)$$

We further define the error exponent of $P_e(\delta)$ as

$$e(\delta) = \lim_{n \to \infty} -\frac{1}{n} \log P_e(\delta). \quad (4)$$

The test $\delta$ is *consistent* if $P_e(\delta)$ converges to zero as $n$ goes to infinity:

$$\lim_{n \to \infty} P_e(\delta) = 0, \quad (5)$$

and the test $\delta$ is *exponentially consistent* if $P_e(\sigma)$ converges to zero exponentially fast with respect to $n$, i.e., $e(\delta) > 0$.

In the existence problem, if the decision is that the testing sequence is generated by one of the existent distributions, then a natural next step is to further identify which distribution generates the testing sequence. We refer to such a problem as the *association problem*, which can be formulated as the following multiple hypothesis testing problem:

$$H_i: p_s = p_i, \text{ for } i = 1, ..., M,$$

where the goal is to determine which hypothesis occurs.

We let $\sigma(\boldsymbol{y})$ denote a test rule for the association problem, which maps the testing sequence $\boldsymbol{y}$ to one of the $M$ hypotheses. Then, for the association problem, we take the following maximum of $M$ error probabilities as the performance metric:

$$P_e(\sigma) = \max_{1 \leqslant i \leqslant M} P(\sigma \neq i | H_i). \quad (6)$$

The error exponent is defined in the same fashion as for the existence problem.

## III. MAIN RESULTS

In this section, we construct the test rules for the existence and association problems and analyze the performances of these test rules. We also discuss computational issues of the error exponents.

### A. Existence Problem: Binary Composite Hypothesis Testing

We first construct a test based on the generalized likelihood. For each known existing distribution, i.e., $p_i$ with $1 \leqslant i \leqslant M_1$, we consider the following likelihood of the testing sequence:

$$\begin{aligned} P(\boldsymbol{y}|H_i) &= \prod_{k=1}^{n} p_i(y_k) \\ &= \exp\left\{-nH(\gamma(\boldsymbol{y})) - nD(\gamma(\boldsymbol{y}) \| p_i)\right\}, \end{aligned} \quad (7)$$

for $1 \leqslant i \leqslant M_1$, where $\gamma(\boldsymbol{y})$ denotes the empirical distribution of $\boldsymbol{y}$ given by

$$\gamma(y) \triangleq \frac{\text{number of samples } y \text{ in } \boldsymbol{y}}{\text{length of } \boldsymbol{y}},$$

$H(\cdot)$ denotes the entropy given by

$$H(p) = -\sum_{y \in \mathcal{Y}} p(y) \log p(y), \quad (8)$$

and $D(\cdot \| \cdot)$ denotes the KL divergence given by

$$D(p\|q) = \sum_{y \in \mathcal{Y}} p(y) \log \frac{p(y)}{q(y)}. \quad (9)$$

Since the term $H(\gamma(\boldsymbol{y}))$ does not depend on the distribution $p_i$, it is dropped from the maximum likelihood test. Hence, the only term that is useful for detection turns out to be

$$D(\gamma(\boldsymbol{y}) \| p_i), \quad (10)$$

which captures the distance between the empirical distribution of the testing sequence and $p_i$. This further suggests that if $p_i$ is unknown, i.e., $i > M_1$, the corresponding likelihood term should still use the form of (10), but using the empirical distribution $\gamma(\boldsymbol{t}_i)$ of $p_i$ obtained from the training sequence $\boldsymbol{t}_i$ to replace $p_i$ in (10), which is hence

$$D(\gamma(\boldsymbol{y}) \| \gamma(\boldsymbol{t}_i)). \quad (11)$$

In order to construct a test, it is natural to determine $H_0$ if the testing sequence is sufficiently far away from all existing distributions, and determine $H_1$ otherwise. Such a test is given by

$$\delta(\boldsymbol{y}): \min\left\{\min_{i \leqslant M_1} D(\gamma(\boldsymbol{y}) \| p_i), \min_{i > M_1} D(\gamma(\boldsymbol{y}) \| \gamma(\boldsymbol{t}_i))\right\} \underset{H_1}{\overset{H_0}{\gtrless}} \alpha, \quad (12)$$

where $\alpha$ is defined in (1) as the lower bound on the minimum Chernoff information.

Note that the left-hand-side of (12) involves the KL divergence; but the threshold in the right-hand-side corresponds to the Chernoff information. To explain this setup intuitively, we assume that there is only one existing distribution ($M = 1$), $p_0$, in the following argument. With an observed $\boldsymbol{y}$ such that $D(\gamma(\boldsymbol{y}) \| p_0) > \alpha$, there exists an alternative distribution $d$, where $C(p_0, d) = \alpha$, satisfying $D(\gamma(\boldsymbol{y}) \| d) < \alpha$. Hence $d$ can generate $\boldsymbol{y}$ with a larger probability than $p_0$ and the decision is $H_0$. With an observed $\boldsymbol{y}$ such that $D(\gamma(\boldsymbol{y}) \| p_0) < \alpha$, we have $D(\gamma(\boldsymbol{y}) \| d) > \alpha$ for any $d$ with $C(p_0, d) \leqslant \alpha$. Hence, $\boldsymbol{y}$ is more likely to be generated by $p_0$ than any alternative distributions and the decision is $H_1$. The reader can refer to [2] for more details.

In order to understand the nonparametric model, we first analyze the parametric model with $M_1 = M$. Since all distributions $p_i$'s are known, test (12) becomes

$$\delta(\boldsymbol{y}): \min_{i} D(\gamma(\boldsymbol{y}) \| p_i) \underset{H_1}{\overset{H_0}{\gtrless}} \alpha. \quad (13)$$

The performance of such a test is summarized in the following lemma.

**Lemma 1.** *Test* (13) *is exponentially consistent and the exponent of the maximum error probability is*

$$e = \min\{e_1, e_2\}, \quad (14)$$

*where $e_1$ is the solution to the following minimization problem:*

$$\min_{q, d \in \Delta} D(q \| d) \quad (15)$$
$$\text{s.t.} \quad C(d, p_i) \geqslant \alpha, \text{ for } i \leqslant M,$$
$$q \in E,$$

*with $\Delta = \{q: \sum_{y \in \mathcal{Y}} q(y) = 1, 0 \leqslant q(y) \leqslant 1\}$ and*

$$E = \{q: \exists i \text{ s.t. } D(q \| p_i) \leqslant \alpha\},$$

and $e_2$ is the solution to the following minimization problem:

$$\min_{j \leqslant M_1} \min_{q \in \Delta} D(q\|p_j) \qquad (16)$$
$$s.t. \quad q \in \overline{E},$$

with $\overline{E}$ being the complementary set of $E$.

Furthermore, the error exponent is lower-bounded by $\alpha$, i.e., $e \geqslant \alpha$.

*Proof.* See Appendix B. □

The above result suggests that the error exponent is lower-bounded by the parameter $\alpha$, which is the distance between the nearest alternative distribution with the set of composite distributions.

We next analyze test (12) for the nonparametric model, in which $p_i$'s for $i > M_1$ are unknown. Clearly, the performance of the test depends on $\frac{\bar{n}}{n}$, i.e., the ratio between the lengths of the training and testing sequences. We let $\beta = \lim_{n \to \infty} \frac{\bar{n}}{n}$. Then the following theorem summarizes the performance of test (12).

**Theorem 1.** *The exponent of the maximum error probability for test* (12) *is*

$$e = \min\{e_1, e_2\}, \qquad (17)$$

*where $e_1$ is the solution to the following minimization problem:*

$$\min_{d,q,q_{M_1+1},\ldots,q_M \in \Delta} D(q\|d) + \beta \sum_{k=M_1+1}^{M} D(q_k\|p_k) \quad (18)$$
$$s.t. \quad C(d, p_i) \geqslant \alpha, \text{ for } i \leqslant M,$$
$$(q, q_{M_1+1}, \ldots, q_M) \in E,$$

*and $e_2$ is the solution to the following minimization problem:*

$$\min_{i \leqslant M} \min_{q,q_{M_1+1},\ldots,q_M \in \Delta} D(q\|p_i) + \beta \sum_{k=M_1+1}^{M} D(q_k\|p_k) \quad (19)$$
$$s.t. \quad (q, q_{M_1+1}, \ldots, q_M) \in \overline{E}.$$

*Here*

$$E = \{(q, q_{M_1+1}, \ldots, q_M) : \exists i \leqslant M_1 \text{ s.t. } D(q\|p_i) \leqslant \alpha$$
$$\text{or } \exists i > M_1 \text{ s.t. } D(q\|q_i) \leqslant \alpha\}, \quad (20)$$

*and $\overline{E}$ is the complementary set of $E$.*

*If $\beta > 0$, then test* (12) *is exponentially consistent. If $\beta = \infty$, then the error exponent is equal to that of the parametric model, and is hence optimal.*

*Proof.* The proof is provided in Section III-D. □

The comparison of Lemma 1 and Theorem 1 implies that the error exponent for the nonparametric model is highly affected by the parameter $\beta$, which is the ratio between the lengths of training and testing sequences. A larger $\beta$ results in a better performance. As long as the length of training sequences is not too short compared with the testing sequence such that $\beta > 0$, the test is exponentially consistent. Furthermore, if $\beta = \infty$, i.e., the length of training sequences scales order-wise faster than that of the testing sequence, the error exponent is equal to that of the parametric model. This is reasonable since the hypothesis distributions are well estimated due to the large sample size, so that the performance achieves that of the parametric case.

**Corollary 1.** *If $\beta = 0$ and $M_1 < M$ (at least one distribution is unknown), the error exponent of test* (12) *is zero, i.e., the test is not exponentially consistent.*

*Proof.* Consider the error exponent $e_1$ given in (18). If $\beta = 0$, $q_M$ can be set such that

$$C(q_M, p_i) > \alpha, \text{ for } i \leqslant M_1 \qquad (21)$$

and

$$C(q_M, q_i) > \alpha, \text{ for } M_1 < i < M. \qquad (22)$$

Then let $q = q_M$ and $d = q_M$. This setting satisfies the constraint, and thus $e_1 = 0$. □

The condition $\beta = 0$ implies that the length of training sequences scales order-wise slower than that of the testing sequence. In such a case, the distributions corresponding to the hypotheses cannot be well estimated, which causes the inconsistency of the test.

*B. Association Problem: Multiple Hypothesis Testing*

The problem of the *parametric* multiple hypothesis testing problem has been well studied in the literature. It has been shown (see [1], [2]) that the following maximum likelihood test achieves the optimal exponent of the maximum error probability:

$$\sigma(\boldsymbol{y}) = \arg\max_i P(\boldsymbol{y}|H_i). \qquad (23)$$

We here focus on the nonparametric case. We construct a generalized maximum likelihood test by replacing each unknown distribution $p_i$ (for $i > M_1$) in (23), corresponding to hypothesis $i$, with the empirical distribution of the training sequence $\boldsymbol{t}_i$. The resulting test is given by

$$\sigma(\boldsymbol{y}) = \arg\max_i \left\{ \begin{array}{ll} P(\boldsymbol{y}|p_i), & \text{if } i \leqslant M_1 \\ P(\boldsymbol{y}|\gamma(\boldsymbol{t}_i)), & \text{if } i > M_1 \end{array} \right\}. \qquad (24)$$

Applying (7), test (24) is equivalent to

$$\sigma(\boldsymbol{y}) = \arg\min_i \left\{ \begin{array}{ll} D(\gamma(\boldsymbol{y})\|p_i), & \text{if } i \leqslant M_1 \\ D(\gamma(\boldsymbol{y})\|\gamma(\boldsymbol{t}_i)), & \text{if } i > M_1 \end{array} \right\}. \qquad (25)$$

The following theorem characterizes the performance of test (25).

**Theorem 2.** *Apply test* (25) *to the nonparametric multiple hypothesis testing problem. The error exponent of the maximum error probability is given by*

$$\min_{i,j: i \neq j} e_{i,j},$$

*where $e_{i,j}$ is given as follows.*
- *For $i \leqslant M_1$ and $j \leqslant M_1$, $e_{i,j} = C(p_i, p_j)$;*
- *For $i \leqslant M_1$ and $j > M_1$,*

$$e_{i,j} = \min_{q,q_j \in \Delta} D(q\|p_j) + \beta D(q_j\|p_j) \qquad (26)$$
$$s.t. \quad D(q\|q_j) \geqslant D(q\|p_i);$$

- *For $i > M_1$ and $j \leqslant M_1$,*
$$e_{i,j} = \min_{q_i \in \Delta} \; C(q_i, p_j) + \beta D(q_i \| p_i); \quad (27)$$

- *For $i > M_1$ and $j > M_1$,*
$$e_{i,j} = \min_{q,q_i,q_j \in \Delta} \; D(q\|p_j) + \beta D(q_i\|p_i) + \beta D(q_j\|p_j) \quad (28)$$
$$\text{s.t.} \quad D(q\|q_j) \geqslant D(q\|q_i).$$

*Proof.* The proof is provided in Section III-D. □

Theorem 2 implies that the error exponent of test (25) is determined by the smallest $e_{i,j}$, which captures the error exponent for the case where the ground truth is $H_j$ but $\sigma(\boldsymbol{y}) = i$. If both $p_i$ and $p_j$ are known, $e_{i,j}$ equals the Chernoff information between $p_i$ and $p_j$. Thus, if $M_1 = M$, i.e., all distributions are known, the error exponent reduces to that of the parametric hypothesis testing problem given in [2]. If $p_i$ is known, but $p_j$ is unknown, (26) consists of two terms: the second term captures the approximation error of the training sequence to $p_j$ (where $q_j$ can be viewed as the approximation of $p_j$), and the first term represents the detection error (where $q$ can be viewed as the approximation of the testing distribution). Hence, if $q_j = p_j$ (i.e., $p_j$ is perfectly learned from the training sequence), $e_{i,j}$ becomes $C(p_i, p_j)$ (the parametric case). This also implies that $e_{i,j}$ in such a case is no larger than $C(p_i, p_j)$. If $p_i$ is unknown but $p_j$ is known, (27) also consists of two terms: the second term captures the approximation error of the training sequence to $p_i$ (where $q_i$ can be viewed as the approximation of $p_i$), and the first term represents the detection error (where $q$ can be viewed as approximation of the testing distribution). If neither $p_i$ nor $p_j$ is known, (28) consists of three terms: the last two terms correspond respectively to the approximation errors of $p_i$ and $p_j$, and the first term represents the detection error. In this case, $e_{i,j}$ reduces to $C(p_i, p_j)$ if $q_i = p_i$ and $q_j = p_j$ (i.e., the approximations of the distributions are perfect). Thus, the error exponent in this case is no larger than $C(p_i, p_j)$.

The following corollary explains under what conditions the error exponent of test (25) approaches that for parametric hypothesis testing, which serves as an upper bound.

**Corollary 2.** *If $\beta > 0$, test (25) is exponentially consistent. Especially, if $\beta \to \infty$, then the error exponent goes to $\min_{\{i,j: i \neq j\}} C(p_i, p_j)$, which is optimal, i.e., the error exponent of the nonparametric case approaches that of the parametric case if the length of training sequences scales order-wise faster than that of the testing sequence.*

*Proof.* Clearly, if $\beta > 0$, the values of (27) is strictly larger than zero. In (26), if $q = p_j = q_j$, we have $D(q\|q_j) = 0$ and $D(q\|p_i) > 0$, which contradicts the constraint. The argument is similar for (28).

If $\beta \to \infty$, then to achieve the minimum value among (26), (27) and (28), $q_i$ and $q_j$ must be set as $p_i$ and $p_j$. Then it is clear that $e_{i,j}$ in all three cases equals $C(p_i, p_j)$. Thus, the error exponent of the problem equals $\min_{\{i,j: i \neq j\}} C(p_i, p_j)$. □

To further explain the above result, if $0 < \beta < \infty$, for a larger $\beta$, the error between $\gamma(\boldsymbol{t}_i)$ and $p_i$ is smaller, and hence the exponent of the maximum error probability takes a larger value. The error exponent is strictly larger than zero. In the extreme case with $\beta = \infty$, the error exponent equals that of the fully parametric model, and hence achieves the optimal value. Thus, if the length of training sequences increases much faster than that the testing sequence, the error between $\gamma(\boldsymbol{t}_i)$ and $p_i$ can be ignored. In such a case, those unknown distributions can be accurately estimated, and hence do not affect the error exponent of the maximum error probability.

**Corollary 3.** *If $\beta = 0$ and $M_1 < M$ (at least one distribution is unknown), then the error exponent of the maximum error probability for test (25) equals zero, i.e., the test is not exponentially consistent.*

*Proof.* Consider (27). If $\beta = 0$, (27) becomes $\min_{q_j \in \Delta} C(p_i, q_j)$, which equals zero with $q_j = p_i$. Similarly, (28) becomes
$$\min_{q,q_i,q_j \in \Delta} \; D(q\|p_j) \quad (29)$$
$$\text{s.t.} \quad D(q\|q_j) \geqslant D(q\|q_i),$$
whose optimal value equals zero with $q = p_j$. Since $M_1 < M$, as least one pair $(i,j)$ satisfies $i \leqslant M_1, j > M_1$, or $i > M_1, j > M_1$. Hence, the error exponent must be zero. □

The above result implies that if the length of training sequences scales order-wise slower than that of the testing sequence, then the unknown distributions cannot be well estimated, which consequently causes the inconsistency of the test.

### C. Computation of Error Exponent

It is clear in our analysis that the error exponent in various cases is expressed as the minimum value of an optimization problem, which is nonconvex and difficult to solve. We next discuss how to obtain solutions to these optimization problems.

First, problem (27) is a min-max problem, which can be written as
$$\min_{q_j \in \Delta} \max_{\lambda \in [0,1]} F(q_j, \lambda) = -\log\left(\sum_{y \in \mathcal{Y}} p_j(y)^\lambda q_i(y)^{1-\lambda}\right) + \beta \sum_{y \in \mathcal{Y}} q_i(y) \log \frac{q_i(y)}{p_j(y)}. \quad (30)$$

It is easy to prove that the objective function is convex over $q_j$ and concave over $\lambda$, and for every saddle point $(\hat{q}_j, \hat{\lambda})$, we have
$$\inf_{q_j \in \Delta} \sup_{\lambda \in [0,1]} F(q_j, \lambda) \leqslant F(\hat{q}_j, \hat{\lambda}) \leqslant \sup_{\lambda \in [0,1]} \inf_{q_j \in \Delta} F(q_j, \lambda). \quad (31)$$

Thus, with the following lemma, all saddle points of problem (30) share the same function value for $F(\cdot, \cdot)$.

**Lemma 2.** (Corollary 37.3.2 in [27]) *Let $C$ and $D$ be nonempty closed convex sets in $R^m$ and $R^n$, respectively, and let*

$K$ be a continuous finite concave-convex function on $C \times D$. If either $C$ or $D$ is bounded, we have

$$\inf_{v \in D} \sup_{u \in C} K(u,v) = \sup_{u \in C} \inf_{v \in D} K(u,v). \tag{32}$$

In addition, following from [27], the optimal point of the min-max problem is one of the saddle points of the objective function. Thus, with the discussion above, the problem is solved at a satisfactory level as long as we find one saddle point. A sub-gradient method [28] can be utilized to find such a point. Namely, first initialize $q_j^{(0)}$ and $\lambda^{(0)}$ randomly, and then perform sub-gradient decent steps alternatively over $q_j$ and $\lambda$ as

$$q_j^{(k+1)} = \mathcal{P}_\Delta[q_j^{(k)} - s\nabla F_{q_j}(q_j^{(k)}, \lambda^{(k)})] \tag{33}$$

$$\lambda^{(k+1)} = \mathcal{P}_{[0,1]}[\lambda^{(k)} + s\nabla F_\lambda(q_j^{(k+1)}, \lambda^{(k)})], \tag{34}$$

where $s$ is the step size, $\mathcal{P}_\Delta$ and $\mathcal{P}_{[0,1]}$ denote the projections onto the simplex sets $\Delta$ and $[0,1]$, respectively; $\nabla F_{q_j}$ and $\nabla F_\lambda$ denote the sub-gradients of $F$ with respect to $q_j$ and $\lambda$, respectively. Then by choosing an appropriate step size $s$, the above algorithm can be shown to converge.

The problem in (26) is also a nonconvex problem since the constraint set is nonconvex. Hence, it is difficult to make the projection onto the constraint set. In this case, we incorporate the constraint set into the objective function as

$$\min_{q,q_j \in \Delta} G(q,q_j) = D(q\|p_j) + \beta D(q_j\|p_j) + l(D(q\|q_j) - D(q\|p_i)), \tag{35}$$

where

$$l(x) = \begin{cases} 0, & x \geqslant 0 \\ \frac{1}{2}\mu x^2, & x < 0 \end{cases} \tag{36}$$

for some $\mu > 0$. To minimize the difference between (26) and (35), we need to set a large value for $\mu$. It can be shown that (26) is a Kurdyka-Lojasiewicz (KL) function [29] and Lipschitz continuous near the critical point. Then, we apply the following gradient projection method [29], [30]

$$q_j^{(k+1)} = \mathcal{P}_\Delta[q_j^{(k)} - s\nabla G_{q_j}(q^{(k)}, q_j^{(k)})] \tag{37}$$

$$q^{(k+1)} = \mathcal{P}_\Delta[q^{(k)} - s\nabla G_q(q^{(k)}, q_j^{(k)})], \tag{38}$$

where $s$ is the step size, $\mathcal{P}_\Delta$ denotes the projection onto the simplex set $\Delta$, and $\nabla G_q$ and $\nabla G_{q_j}$ denote the sub-gradients of $G$ with respect to $q$ and $q_j$, respectively. By choosing a $q_j^{(0)}$ to be close to $p_j$ (e.g., let $q_j^{(0)}$ takes the empirical distribution of $t_j$), and $q^{(0)}$ be in the middle of $q_j^{(0)}$ and $p_i$, the iteration can be shown to converge to a local minimizer of (35).

Problem (28) can be computed similarly as (26), which can be converted to

$$\min_{q,q_i,q_j \in \Delta} H(q,q_i,q_j) = D(q\|p_j) + \beta D(q_i\|p_i) + \beta D(q_j\|p_j) + l(D(q\|q_j) - D(q\|q_i)). \tag{39}$$

We then take the gradient projection method as

$$q_i^{(k+1)} = \mathcal{P}_\Delta[q_i^{(k)} - s\nabla H_{q_i}(q^{(k)}, q_i^{(k)}, q_j^{(k)})] \tag{40}$$

$$q_j^{(k+1)} = \mathcal{P}_\Delta[q_j^{(k)} - s\nabla H_{q_j}(q^{(k)}, q_i^{(k)}, q_j^{(k)})] \tag{41}$$

$$q^{(k+1)} = \mathcal{P}_\Delta[q^{(k)} - s\nabla H_q(q^{(k)}, q_i^{(k)}, q_j^{(k)})]. \tag{42}$$

By choosing a $q_i^{(0)}$ near $p_i$, $q_j^{(0)}$ near $p_j$, and $q^{(0)}$ in the middle of $q_j^{(0)}$ and $q_i^{(0)}$, the iteration converges to a local minimizer of (39).

### D. Technical Proofs

In this subsection, we provide the proofs of Theorems 1 and 2.

*Proof of Theorem 1.* Following test (12), the maximum error probability is given by

$$P_e(\delta) = \max\{P(\delta = 1|H_0), P(\delta = 0|H_1)\}. \tag{43}$$

The type I error probability is given by

$$\begin{aligned}&P(\delta = 1|H_0)\\ =&P\big(\min\{\min_{i \leqslant M_1} D(\gamma(\boldsymbol{y})\|p_i),\\ &\min_{i > M_1} D(\gamma(\boldsymbol{y})\|\gamma(\boldsymbol{t}_i))\} \leqslant \alpha \mid H_0\big)\\ =&P\big((\gamma(\boldsymbol{y}), \gamma(\boldsymbol{t}_{M_1+1}), ..., \gamma(\boldsymbol{t}_M)) \in E \mid H_0\big),\end{aligned}$$

where $E$ is defined in (20) and $H_0$ denotes the hypothesis that $\boldsymbol{y}$ is not generated by one of the $M$ distributions, i.e., $\{\boldsymbol{t}_k \sim p_k,$ for $k > M_1;$ $\boldsymbol{y} \sim d$ with $C(d, p_i) \geqslant \alpha,$ for $i = 1, ..., M\}$. Then the exponent of $P(H_1|H_0)$ follows from Lemma 3 as the generalization of the Sanov's theorem in Appendix A.

Similarly, the type II error probability is given by

$$\begin{aligned}&P(\delta = 0|H_1)\\ =&\max_j P\big(\min\{\min_{i \leqslant M_1} D(\gamma(\boldsymbol{y})\|p_i),\\ &\min_{i > M_1} D(\gamma(\boldsymbol{y})\|\gamma(\boldsymbol{t}_i))\} \geqslant \alpha \mid H_{1,j}\big)\\ =&\max_j P\big((\gamma(\boldsymbol{y}), \gamma(\boldsymbol{t}_{M_1+1}), ..., \gamma(\boldsymbol{t}_M)) \in \overline{E} \mid H_{1,j}\big),\end{aligned}$$

where $\overline{E}$ is the complementary set of $E$ and $H_{1,j}$ denotes the hypothesis that $\boldsymbol{y}$ is generated by $p_j$, i.e., $\{\boldsymbol{t}_k \sim p_k,$ for $k > M_1;$ $\boldsymbol{y} \sim p_j\}$. Then the exponent of $P(H_0|H_1)$ also follows from Lemma 3 as the generalization of the Sanov's theorem in Appendix A.

We next analyze the error exponent. For type I error, the error exponent is

$$e_1 := \lim_{n \to \infty} \frac{1}{n} \log P(\delta = 1|H_0). \tag{44}$$

We compute the value of $e_1$ by Lemma 3 as follows

$$\min_{d,q,q_{M_1+1},...,q_M \in \Delta} D(q\|d) + \beta \sum_{k=M_1+1}^{M} D(q_k\|p_k) \tag{45}$$

$$s.t. \quad C(d, p_i) \geqslant \alpha, \text{ for } i \leqslant M,$$

$$(q, q_{M_1+1}, ..., q_M) \in E.$$

Since $E$ in (20) is equal to a union of sets given by

$$E = \{(q, q_{M_1+1}, ..., q_M) : D(q\|p_1) \leqslant \alpha\} \tag{46}$$

$$......$$

$$\cup \{(q, q_{M_1+1}, ..., q_M) : D(q\|p_{M_1}) \leqslant \alpha\}$$

$$\cup \{(q, q_{M_1+1}, ..., q_M) : D(q\|q_{M_1+1}) \leqslant \alpha\}$$

$$......$$

$$\cup \{(q, q_{M_1+1}, ..., q_M) : D(q\|q_M) \leqslant \alpha\},$$

$e_1$ equals
$$\min_j r_{I,j}, \quad (47)$$
where $r_{I,j}$ is given as follows.

If $j > M_1$, then $r_{I,j}$ is the optimal value of
$$\min_{d,q,q_j \in \Delta} D(q\|d) + \beta D(q_j\|p_j) \quad (48)$$
$$\text{s.t.} \quad D(q\|q_j) \leq \alpha,$$
$$C(d,p_i) \geq \alpha, \text{ for } i \leq M.$$

When $0 < \beta < \infty$, the solution is strictly larger than zero. Otherwise, the objective function equals zero. It means that $q = d$ and $q_i = p_i$ hold, and then $C(q,q_j) \geq \alpha$, which contradicts $D(q\|q_j) \leq \alpha$. If $\beta = \infty$ and $D(q_j\|p_j) = 0$, then (48) is equivalent to (49). As a conclusion, the exponent $e_1$ is strictly larger than zero.

If $j \leq M_1$, then $r_{I,j}$ is the optimal value of
$$\min_{d,q \in \Delta} D(q\|d) \quad (49)$$
$$\text{s.t.} \quad D(q\|p_j) \leq \alpha,$$
$$C(d,p_i) \geq \alpha, \text{ for } i \leq M.$$

Problem (49) is similar to (75), which is proved to be no less than $\alpha$ in Appendix B. Thus $r_{I,j} \geq \alpha$ for $j \leq M_1$.

For the type II error, the error exponent is given by
$$e_2 := \lim_{n \to \infty} \frac{1}{n} \log P(\delta = 0 | H_1). \quad (50)$$

We compute the value of $e_2$ by Lemma 3 as follows
$$\min_{i \leq M} \min_{q,q_{M_1+1},\ldots,q_M \in \Delta} D(q\|p_i) + \beta \sum_{k=M_1+1}^{M} D(q_k\|p_k) \quad (51)$$
$$\text{s.t.} \quad (q, q_{M_1+1}, \ldots, q_M) \in \overline{E}.$$

If $0 < \beta < \infty$, the optimal value is obviously strictly larger than zero. Otherwise, $q = p_i$ and $q_k = p_k$ for $k > M_1$. It contradicts the constraint $(q, q_{M_1+1}, \ldots, q_M) \in \overline{E}$. If $\beta = \infty$, $D(q_k\|p_k) = 0$ must hold for every $k$. Hence, $e_2$ is the optimal value of
$$\min_{i \leq M} \min_{q \in \Delta} D(q\|p_i) \quad (52)$$
$$\text{s.t.} \quad D(q\|p_k) \geq \alpha, \text{ for } k = 1, \ldots, M,$$

which is $\alpha$. $\square$

*Proof of Theorem 2.* We first define $\bar{p}_i$ such that $\bar{p}_i = p_i$ for $i \leq M_1$, and $\bar{p}_i = \gamma(\boldsymbol{t}_i)$ for $i > M_1$.

The maximum error probability is given by
$$P_e(\sigma) = \max_i P(\sigma \neq i | H_i)$$
$$= \max_i P\{\exists k, D(\gamma(\boldsymbol{y})\|\bar{p}_k) \leq D(\gamma(\boldsymbol{y})\|\bar{p}_i)|H_i\}.$$

It is easy to see that
$$\max_{j \neq i} P\{D(\gamma(\boldsymbol{y})\|\bar{p}_j) \leq D(\gamma(\boldsymbol{y})\|\bar{p}_i)|H_i\}$$
$$\leq P\{\exists k, D(\gamma(\boldsymbol{y})\|\bar{p}_k) \leq D(\gamma(\boldsymbol{y})\|\bar{p}_i)|H_i\}$$
$$\leq \sum_{j \neq i} P\{D(\gamma(\boldsymbol{y})\|\bar{p}_j) \leq D(\gamma(\boldsymbol{y})\|\bar{p}_i)|H_i\}. \quad (53)$$

The exponent of the right-hand-side of (53) can be bounded as
$$\lim_{n \to \infty} -\frac{1}{n} \log \sum_{j \neq i} P\{D(\gamma(\boldsymbol{y})\|r_j) \leq D(\gamma(\boldsymbol{y})\|r_i)|H_i\}$$
$$\leq \lim_{n \to \infty} -\frac{1}{n} \log M \max_{j \neq i} P\{D(\gamma(\boldsymbol{y})\|r_j) \leq D(\gamma(\boldsymbol{y})\|r_i)|H_i\},$$
$$= \lim_{n \to \infty} -\frac{1}{n} \log \max_{j \neq i} P\{D(\gamma(\boldsymbol{y})\|r_j) \leq D(\gamma(\boldsymbol{y})\|r_i)|H_i\},$$
$$\quad (54)$$

where the last equality is because $M$ is a fixed constant. Hence, the exponent of the right-hand-side of (53) is equal to that of the left-hand-side of (53).

Thus, the error exponent of the maximum error probability is given by
$$E(\sigma) = \lim_{n \to \infty} -\frac{1}{n} \log P_e(\sigma) \quad (55)$$
$$= \lim_{n \to \infty} -\frac{1}{n} \log \max_i P(\sigma \neq H_i|H_i) \quad (56)$$
$$= \min_i \lim_{n \to \infty} -\frac{1}{n} \log \max_{j \neq i} P\{D(\gamma(\boldsymbol{y})\|\bar{p}_j)$$
$$\leq D(\gamma(\boldsymbol{y})\|\bar{p}_i)|H_i\} \quad (57)$$
$$= \min_{i,j: i \neq j} \lim_{n \to \infty} -\frac{1}{n} \log P\{D(\gamma(\boldsymbol{y})\|\bar{p}_j)$$
$$\leq D(\gamma(\boldsymbol{y})\|\bar{p}_i)|H_i\}. \quad (58)$$

Let
$$e_{i,j} := \lim_{n \to \infty} -\frac{1}{n} \log P\{D(\gamma(\boldsymbol{y})\|\bar{p}_j) \leq D(\gamma(\boldsymbol{y})\|\bar{p}_i)|H_i\}, \quad (59)$$
which represents the exponent of the probability for event that $\boldsymbol{y}$ is generated by $p_i$ but the GL-test is more likely decide $H_j$ than $H_i$. More formally, $H_i$ denotes the event $\{\boldsymbol{t}_k \sim p_k, \text{ for } k > M_1; \boldsymbol{y} \sim p_j\}$. We next provide the forms for $e_{i,j}$ in four cases.

If $i \leq M_1$ and $j \leq M_1$, then $\bar{p}_i = p_i$ and $\bar{p}_j = p_j$, which are known distributions. Define $E = \{p : D(p, p_j) \leq D(p, p_i)\}$. Then, following the Sanov's theorem in [2], we obtain
$$e_{i,j} = \min_{q \in \Delta} D(q\|p_i) \quad (60)$$
$$\text{s.t.} \quad D(q\|p_j) \leq D(q\|p_i)$$
$$= C(p_i, p_j),$$
where the last equality follows by solving the Lagrangian dual of the optimization problem.

If $i \leq M_1$ and $j > M_1$, then $\bar{p}_i = p_i$ which is known; but $\bar{p}_j = \gamma(\boldsymbol{t}_j)$ is the empirical distribution of the training sequence $\boldsymbol{t}_j$. Following from Lemma 3 in Appendix A, we obtain
$$e_{i,j} = \min_{q,q_j \in \Delta} \lim_{n \to \infty} D(q\|p_j) + \beta D(q_j\|p_j)$$
$$\text{s.t.} \quad D(q\|q_j) \geq D(q\|p_i).$$

If $i > M_1$ and $j \leqslant M_1$, then $\bar{p}_j = p_j$, which is known; but $\bar{p}_i = \gamma(\boldsymbol{t}_i)$, which is the empirical distribution of the training sequence. Following from Lemma 3 in Appendix A, we obtain

$$\begin{aligned} e_{i,j} &= \min_{q, q_i \in \Delta} \quad D(q\|p_j) + \beta D(q_i\|p_i) \\ & \quad s.t. \quad D(q\|p_j) \geqslant D(q\|q_i) \\ &= \min_{q_i \in \Delta} \quad \beta D(q_i\|p_i) + \begin{cases} \min_q & D(q\|p_j) \\ s.t. & D(q\|p_j) \geqslant D(q\|q_i) \end{cases} \\ &= \min_{q_i \in \Delta} \quad C(q_i, p_j) + \beta D(q_i\|p_i), \end{aligned} \quad (61)$$

where the last step follows since the inner optimization is the same as (60).

If $i > M_1$ and $j > M_1$, then $\bar{p}_i = \gamma(\boldsymbol{t}_i)$ and $\bar{p}_j = \gamma(\boldsymbol{t}_j)$, both of which are the empirical distributions of the corresponding training sequences. Following from the Sanov's theorem, we obtain

$$\begin{aligned} e_{i,j} &= \min_{q, q_i, q_j \in \Delta} \quad D(q\|p_j) + \beta D(q_i\|p_i) + \beta D(q_j\|p_j) \\ & \quad s.t. \quad D(q\|q_j) \geqslant D(q\|q_i). \end{aligned}$$

□

## IV. NUMERICAL RESULTS

In this section, we provide numerical results to validate the theoretical analyses for both the existence problem and the association problem.

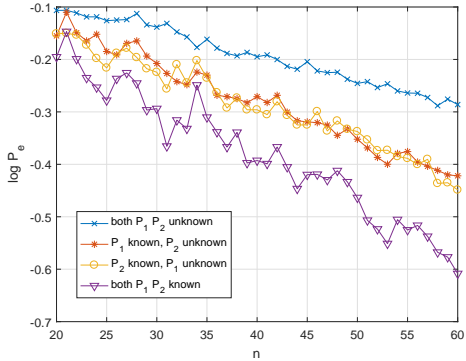

Fig. 1. Existence problem: impact of knowledge of distributions on error decay performance

We first study the existence problem, i.e., the binary composite hypothesis testing. For each specific experiment setting, we repeat 5000 random tests for each of the two cases with $\boldsymbol{y}$ respectively created by one of the existing distributions or generated by a distribution $d$, where $C(d, p_i) \geqslant \alpha$ for every $p_i$ in the given set of composite distributions. In our experiments, $d$ is randomly generated over the support set with probability mass being sampled uniformly from $[0, 1]$ followed by a normalization. Then, bad $d$'s are rejected to make sure $C(d, p_i) \geqslant \alpha$ for every $p_i$.

In the the first experiment, we set $M = 2$, $p_1 = [0.1, 0.1.0.8]$, $p_2 = [0.8, 0.1, 0.1]$, $\alpha = 0.01$, and $\bar{n} = n$, i.e., the length of the training sequences is the same as that of the testing sequence. Fig. 1 plots the changes of $\log(P_e)$ with $n$ for four cases with each distributions known or unknown, respectively. It is clear that the parametric case with both distributions known has the best error decay, and the nonparametric case with neither distribution known has the worst error decay. Furthermore, the approximate linearity of the curves implies exponential decay behaviors for all cases, where the slope $\frac{1}{n}\log(P_e)$ of each curve approximates the error exponent. We then study how the number of composite distributions affects the performance of the GL test (7). We consider three distributions: $p_1 = [0.8, 0.1, 0.1]$, $p_2 = [0.1, 0.1, 0.8]$, $p_3 = [0.6, 0.2, 0.2]$, $\alpha = 0.01$, $\bar{n} = n$, and $M_1 = 0$, i.e., all distributions are assumed unknown. We conduct the experiment for three cases: the set of composite distributions consists of only $p_1$, consists of $p_1$ and $p_2$, and consists of all three distributions. Fig. 2 plots the changes of $\log P_e$ with $n$ for the above cases. It can be seen that the number of composite distributions does not significantly affect the performance in term of the decaying speed. In fact, for the existence problem, the distance between the alternative distribution $d$ and the composite set (i.e., the threshold $\alpha$) is the major parameter that affects the performance. The number of composite distributions should not affect the performance substantially as long as all composite distributions are far away from the alternative distribution $d$.

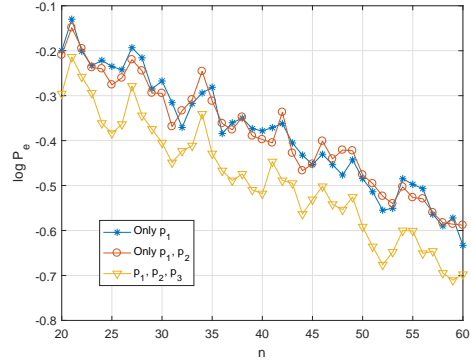

Fig. 2. Existence problem: impact of the number of composite distributions on error decay performance

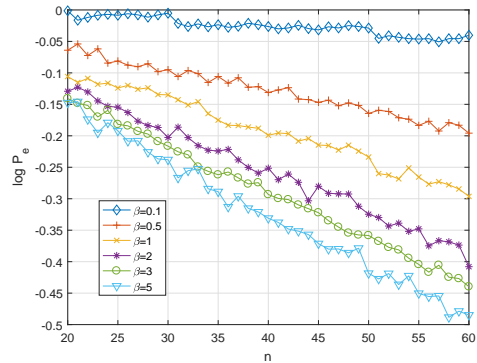

Fig. 3. Existence problem: impact of the length of training sequences on error decay performance

We next study how the ratio $\beta = \frac{\bar{n}}{n}$ affects the performance. We set $p_1 = [0.3, 0.2, 0.5]$, $p_2 = [0.7, 0.2, 0.1]$, $\bar{n} = n$, $\alpha = 0.01$ and $M_1 = 0$. Fig. 3 plots the changes of $\log P_e$ with the length $n$ of the sequences for the cases with $\beta = 0.1, 0.5, 1, 2, 3, 5$, respectively. It can be seen that a larger $\beta$ results in a larger error exponent, which corroborates our theorem. We then study how the distance $\alpha$ between the composite distributions and the alternative distribution affects the test performance. We set $p_1 = [0.3, 0.2, 0.5]$, $p_2 = [0.7, 0.2, 0.1]$, $\bar{n} = n$, and $M_1 = 0$. Fig. 4 plots the changes of $\log P_e$ with $n$ for the cases with $\alpha = 0.005, 0.01, 0.05$, respectively. It can be seen that a larger $\alpha$ results a larger error exponent, which corroborates our theorem. In the next several experiments, we study the association problem, i.e., the multiple hypothesis testing. For each specific experiment setting, we repeat 10000 tests to compute the error probability. We first study how the knowledge of the distributions affects the error decay performance. We set the same experiment parameters as that of the first experiment for the existence problem. Fig. 5 plots the performance for the four cases. Similarly as the existence problem, the parametric case has the best error exponent and the nonparametric case has the worst error exponent.

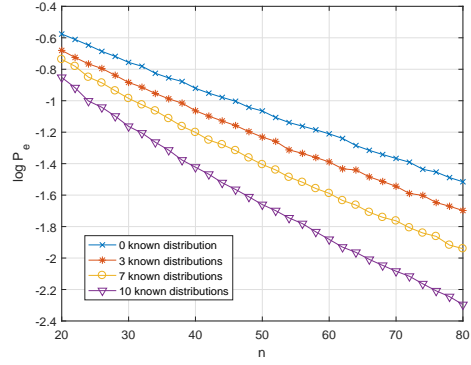

Fig. 6. Association problem: impact of the number of unknown distributions on error decay performance

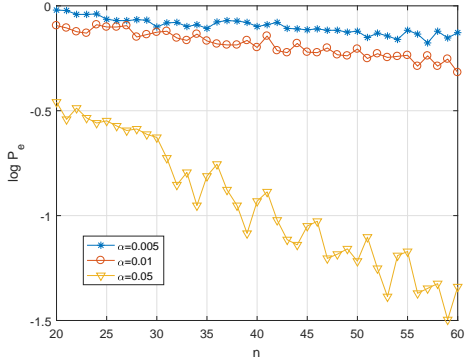

Fig. 4. Existence problem: impact of the number of composite distributions on error decay performance

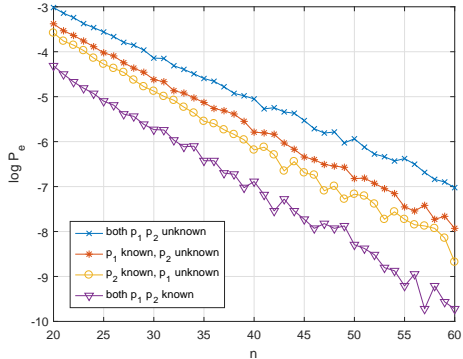

Fig. 5. Association problem: impact of the knowledge of distributions on error decay performance

We then study how the number of unknown distributions affects the error decay performance of test (25). In this experiment, we set $M = 10$, $\bar{n} = n$, and all the 10 distributions are created randomly with support size 3. We study the cases with $M_1 = 0, 3, 7, 10$, respectively. Fig. 6 plots the performance for the four cases. It can be seen that with more distributions known, the error exponent becomes larger, and hence the error decays faster.

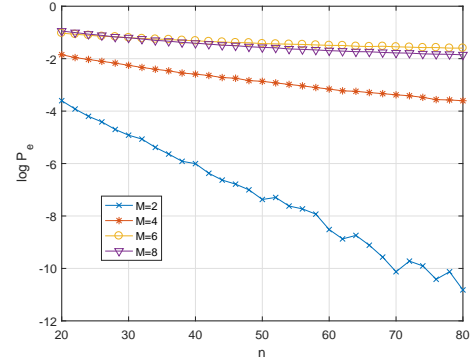

Fig. 7. Association problem: impact of the number of distributions (i.e., hypotheses) on error decay performance

We next study how the number of distributions (i.e., hypotheses) affects the error decay performance. We study the cases that $M = 2, 4, 6, 8$ with $M_1 = \frac{1}{2}M$. All distributions are generated randomly with support size 3 and we set $\bar{n} = n$. Fig. 7 plots the performance for all cases, and it can be seen that a larger $M$ yields a worse performance (i.e., smaller error exponent). This is because the error exponent is determined by the smallest distance among all pairs of distributions, and having more distributions may reduce such smallest distance. We then study how the Chernoff information affects the error decay performance. In this experiment, we set $M = 2$, $M_1 = 0$, $\bar{n} = n$, and study the following three cases. We set $p_1 = [0.9, 0.05, 0.05]$ and $p_2 = [0.05, 0.05, 0.9]$ for Case 1, where $C(p_1, p_2) = 0.746$; $p_1 = [0.8, 0.1, 0.1]$ and $p_2 = [0.1, 0.1, 0.8]$ for Case 2, where $C(p_1, p_2) = 0.4069$; $p_1 = [0.6, 0.2, 0.2]$ and $p_2 = [0.2, 0.2, 0.6]$ for Case 3, where $C(p_1, p_2) = 0.1134$. Fig. 8 plots the performance for all cases, and it can be seen that larger Chernoff distances result in larger error exponents, which corroborates our result.

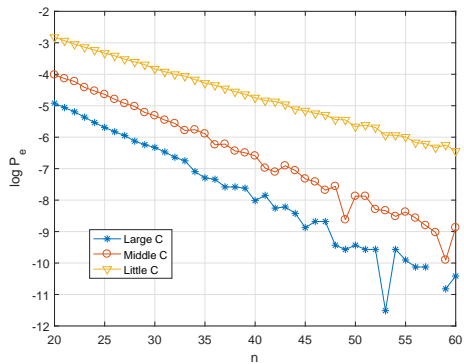

Fig. 8. Association problem: impact of the Chernoff information on error decay performance

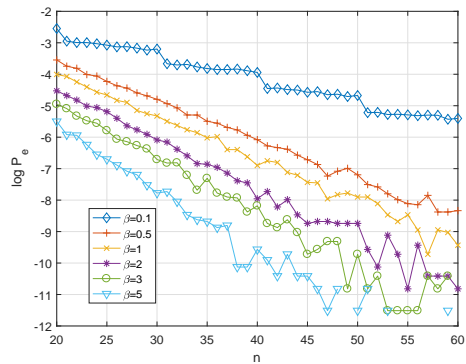

Fig. 10. Association problem: impact of the ratio $\beta = \frac{\bar{n}}{n}$ of the training and testing sequences on error decay performance

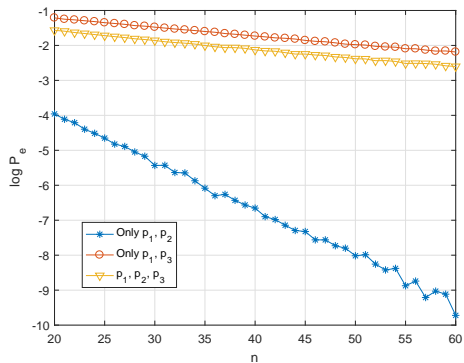

Fig. 9. Association problem: impact of the minimum Chernoff information on error decay performance

We next study how the minimum Chernoff information affects the error decay performance. We consider three distributions: $p_1 = [0.8, 0.1, 0.1]$, $p_2 = [0.1, 0.1, 0.8]$, and $p_3 = [0.6, 0.2, 0.2]$. We study the following three cases. In Case 1, the hypothesis testing is between $p_1$ and $p_2$, where $C(p_1, p_2) = 0.4069$. In Case 2, the hypothesis testing is between $p_1$ and $p_3$, where $C(p_1, p_3) = 0.0247$. In Case 3, the hypothesis testing is among $p_1$, $p_2$ and $p_3$, where the minimum Chernoff information is given by $C(p_1, p_3) = 0.0247$. All distributions are assumed to be unknown and $\bar{n} = n$. Fig. 9 plots the performance for the three cases, and it can be seen that case 1 has the largest error exponent due to its largest minimum Chernoff distance. Furthermore, although case 3 has one more distribution in the hypothesis testing compared to case 2, its error performance is similar to that of case 2 since both cases have the same minimum Chernoff distance. This is in contrast to the existence problem, in which the distance between the composite distributions and the alternative distribution determines the performance.

We finally study how the ratio $\beta = \frac{\bar{n}}{n}$ affects the error decay performance. We set $p_1 = [0.1, 0.1, 0.8]$, $p_2 = [0.8, 0.1, 0.1]$, and $M_1 = 0$. Fig. 10 plots the error decay performance for the cases respectively with $\beta = 0.1, 0.5, 1, 2, 3, 5$, respectively. It can be seen that a larger $\beta$ yields a larger error exponent for test (25), which corroborates Corollary 2.

## V. CONCLUSION

In this paper, we have studied the existence and association problems. Our focus has been on the characterization of the error exponent for the maximum error probability. We have showed that the GL tests are exponentially consistent as long as the number of training samples in each sequence scales no slower than the number of testing samples. As future work, it will be interesting to study the regime where the number of composite distributions in the existence problem or the number of hypotheses in the association problem also goes to infinity, and explore how the number of samples should scale accordingly in order to guarantee the exponential consistency of the GL tests.

## APPENDIX A
## PROOF OF AN EXTENSION OF SANOV'S THEOREM

In this section, we provide an extension of the Sanov's Theorem in [2], which is useful for our proofs.

**Lemma 3.** *Consider a set $\mathcal{P}$ of $M + 1$ distinct distributions $P, P_1, ..., P_M$. Suppose $\boldsymbol{t}$ is a sequence with $n$ i.i.d. samples generated by $P$, and $\boldsymbol{t}_i$ is a sequence with $\bar{n}$ i.i.d. samples generated by $P_i$, for $i = 1, ..., M$, where $\bar{n}$ is a function of $n$. Define $\beta = \lim_{n \to \infty} \frac{\bar{n}}{n}$. Let $E \in \mathcal{P}^{M+1}$ be a set of vectors of distributions. Assume $E$ is the closure of its interior and $\bar{n} = o(e^n)$. Then we have*

$$\lim_{n \to \infty} -\frac{1}{n} P\Big\{ \boldsymbol{t}, \boldsymbol{t}_1, ..., \boldsymbol{t}_M : (\gamma(\boldsymbol{t}), \gamma(\boldsymbol{t}_1), ..., \gamma(\boldsymbol{t}_M)) \in E \Big\}$$
$$= \min_{(p, p_1, ..., p_M) \in E} D(p \| P) + \beta \cdot \sum_i D(p_i \| P_i). \quad (62)$$

*Proof.* The proof follows that of the Sanov's theorem in [2]. For a distribution $P$, let $T(P)$ denote the set of sequences, the empirical distributions of which are the same as $P$, i.e., $T(P)$ denotes the type of $P$ [2].

We first derive an upper bound:

$$P\{\boldsymbol{t},\boldsymbol{t}_1,...,\boldsymbol{t}_M:(\gamma(\boldsymbol{t}),\gamma(\boldsymbol{t}_1),...,\gamma(\boldsymbol{t}_M))\in E\} \quad (63)$$

$$= \sum_{(p,p_1,...,p_M)\in E} P(T(p))\prod_i P_i(T(p_i)) \quad (64)$$

$$\leqslant \sum_{(p,p_1,...,p_M)\in E} \exp\big(-nD(p\|P) - \sum_i \bar{n}D(p_i\|P_i)\big) \quad (65)$$

$$\leqslant \sum_{(p,p_1,...,p_M)\in E} \max_{(p,p_1,...,p_M)\in E} \exp\big(-nD(p\|P) - \sum_i \bar{n}D(p_i\|P_i)\big) \quad (66)$$

$$\leqslant (n+M\bar{n}+1)^{|\mathcal{X}|}\exp\Big(-\min_{(p,p_1,...,p_M)\in E}\big(nD(p\|P) + \sum_i \bar{n}D(p_i\|P_i)\big)\Big), \quad (67)$$

where $|\mathcal{X}|$ is the size of the support set of $\mathcal{P}$.

Considering $\bar{n} = o(e^n)$, we have

$$\lim_{n\to\infty}\frac{1}{n}\log P\{\boldsymbol{t},\boldsymbol{t}_1,...,\boldsymbol{t}_M:(\gamma(\boldsymbol{t}),\gamma(\boldsymbol{t}_1),...,\gamma(\boldsymbol{t}_M))\in E\}$$
$$\leqslant \min_{(p,p_1,...,p_M)\in E} D(p\|P) + \beta\sum_i D(p_i\|P_i).$$

We next derive a lower bound. Since $E$ is a close set, we can find a sequence of distributions $(p^{(n)}, p_1^{(n)}, ..., p_M^{(n)})$ such that $(p^{(n)}, p_1^{(n)}, ..., p_M^{(n)}) \in E \cap \mathcal{P}_{n,\bar{n},...,\bar{n}}$ and

$$\lim_{n\to\infty} D(p^{(n)}\|P) + \frac{\bar{n}}{n}\sum_i D(p_i^{(n)}\|P_i)$$
$$= D(p\|P) + \beta\sum_i D(p_i\|P_i), \quad (68)$$

where $\mathcal{P}_n$ denotes the set of distributions such that the probability of every element is $\frac{k}{n}$ for some integer $k$.

Now for any sufficiently large $n$, we have

$$P\{\boldsymbol{t},\boldsymbol{t}_1,...,\boldsymbol{t}_m:(\gamma(\boldsymbol{t}),\gamma(\boldsymbol{t}_1),...,\gamma(\boldsymbol{t}_M))\in E\}$$
$$= \sum_{(p,p_1,...,p_M)\in E} P(T(p))\prod_i P_i(T(p_i))$$
$$\geqslant P(T(p^{(n)}))\prod_i P_i(T(p_i^{(n)}))$$
$$\geqslant \frac{1}{((n+1)(\bar{n}+1)^M)^{|\mathcal{X}|}}\exp\Big(-nD(p^{(n)}\|P) - \sum_i \bar{n}D(p_i^{(\bar{n})}\|P_i)\Big). \quad (69)$$

Then

$$\lim_{n\to\infty} -\frac{1}{n}\log P\{\boldsymbol{t},\boldsymbol{t}_1,...,\boldsymbol{t}_M:(\gamma(\boldsymbol{t}),\gamma(\boldsymbol{t}_1),...,\gamma(\boldsymbol{t}_M))\in E\}$$
$$\geqslant \min_{(p,p_1,...,p_M)\in E} D(p\|P) + \beta\sum_i D(p_i\|P_i).$$

Combining the upper and lower bounds establishes the lemma. □

## APPENDIX B
## PROOF OF LEMMA 1

We separately bound the two terms in the maximum error probability

$$P_e = \max\{P(\delta=1|H_0), P(\delta=0|H_1)\}. \quad (70)$$

For the test given in (12), we derive

$$P(\delta=1|H_0) = P\big(\min_i D(\gamma(\boldsymbol{y})\|p_i)\leqslant\alpha\mid H_0\big)$$
$$= P\big(\gamma(\boldsymbol{y})\in E\mid H_0\big), \quad (71)$$

where $H_0$ denotes the hypothesis that $y$ is not generated by one of the $M$ distributions, i.e., $\{\boldsymbol{y}\sim d$ with $C(p_i,d)\geqslant\alpha$, for $i=1,...,M\}$, and the set $E$ is given by

$$E = \{q:\exists i\text{ s.t. }D(q\|p_i)\leqslant\alpha\}.$$

Applying the Sanov's theorem, the error exponent of $P(H_1|H_0)$ is given by

$$e_1 = \min_{d,q\in E} D(q\|d) \quad (72)$$
$$\text{s.t.}\quad C(d,p_i)\geqslant\alpha,\quad\text{for }i\leqslant M.$$

We note that the constraint set $E$ can be further written as the union of $m$ subsets as

$$E = \{q:D(q\|p_1)\leqslant\alpha\}$$
$$\cup\{q:D(q\|p_2)\leqslant\alpha\}$$
$$...$$
$$\cup\{q:D(q\|p_M)\leqslant\alpha\}. \quad (73)$$

Thus, the optimization problem (72) is equivalent to finding the optimal solution over each subset and then taking the minimum among the solutions as

$$e_1 = \min_i r_i, \quad (74)$$

where $r_i$ is the solution of

$$\min_q D(q\|d) \quad (75)$$
$$\text{s.t.}\quad D(q\|p_i)\leqslant\alpha,$$
$$C(d,p_i)\geqslant\alpha,\quad\text{for }i\leqslant M.$$

We note that $C(p_i,d)$ is the optimal value of the following optimization problem:

$$\min_{q'} D(q'\|d) \quad (76)$$
$$\text{s.t.}\quad D(q'\|p_i)\leqslant D(q'\|d),$$

or equivalently

$$\min_{q''} D(q''\|p_i) \quad (77)$$
$$\text{s.t.}\quad D(q''\|p_i)\geqslant D(q''\|d).$$

We next prove by contradiction that if $D(q\|p_i)\leqslant C(d,p_i)$, then $D(q\|d)\geqslant C(d,p_i)$ for any distributions $d$ and $p_i$'s. If $D(q\|d) < C(d,p_i)$, then $q$ does not belong to the constraint set (76). Hence, $D(q\|p_i) > D(q\|d)$. As mentioned before, (77) takes the optimal value only when $D(q''\|p_i) = D(q''\|d)$. Hence, $D(q\|p_i) > D(q\|d)$ implies $D(q\|p_i) > C(d,p_i)$, which contradicts the assumption. Thus, the original claim is true.

Now since $C(d, p_i) \geqslant \alpha$ for all $i$ due to the assumption of the problem, the constraint in (75) implies $D(q\|p_i) \leqslant C(d, p_i)$. Due to the above claim, we conclude that $r_i \geqslant C(d, p_i) \geqslant \alpha$ for all $i$. Thus, the error exponent has $e_1 \geqslant \alpha$.

We next derive the error exponent $e_2$ of $P(\delta = 0|H_1)$ and show that it is no less than $\alpha$.

$$P(\delta = 0|H_1) = \max_j P\big(\min_i D(\gamma(\boldsymbol{y})\|p_i) \geqslant \alpha \mid H_{1,j}\big)$$
$$= \max_j P\big(\gamma(\boldsymbol{y}) \in \overline{E} \mid H_{1,j}\big), \quad (78)$$

where $H_{1,j}$ denote the sub-hypothesis that $\boldsymbol{y}$ is generated by $p_j$, i.e., $\{\boldsymbol{y} \sim p_j\}$, and $\overline{E}$ is the complementary set of $E$ given by

$$\overline{E} = \{q : \min_i D(q\|p_i) \geqslant \alpha\}. \quad (79)$$

Applying Sanov's theorem, we obtain the error exponent of $P\big(\gamma(\boldsymbol{y}) \in \overline{E} \mid H_{1,j}\big)$ is

$$e_{2,j} = \min_{q \in \overline{E}} D(q\|p_j), \quad (80)$$

which is clearly no less than $\alpha$ due to the definition of $\overline{E}$.

Thus, the error exponent of $P(\delta = 0|H_1)$ is

$$\min_j e_{2,j}, \quad (81)$$

which is also no less than $\alpha$.

## Acknowledgment

Y. Liu would like to thank Dr. Zhi Ding for his support of Yixian Liu's visit to Syracuse University, where this work was performed.